\documentclass[%
 aip,
cp,  
 amsmath,amssymb,
 reprint,%
]{revtex4-2}

\usepackage{graphicx}
\usepackage{dcolumn}
\usepackage{bm}
\usepackage{cleveref}
\usepackage{xcolor}

\usepackage[utf8]{inputenc}
\usepackage[T1]{fontenc}
\usepackage{mathptmx}

\newcommand{\qbar}{\overline{q}}
\newcommand{\ubar}{\overline{u}}
\newcommand{\dbar}{\overline{d}}
\newcommand{\sbar}{\overline{s}}

\newcommand{\covLap}{\widetilde{\Delta}}

\newcommand{\bra}[1]{\langle#1|}
\newcommand{\ket}[1]{|#1\rangle}

\newcommand{\vacexp}[1]{\langle0|#1|0\rangle}

\DeclareMathOperator{\Tr}{Tr}

\renewcommand{\Re}{\operatorname{Re}}
\renewcommand{\Im}{\operatorname{Im}}

\newcommand{\ie}{i.e.~}

\makeatletter
\newlength\CoolS@sizex
\newlength\CoolS@sizey
\newcommand*\CoolS@inner{%
  \begin{tikzpicture}[baseline=0.04\CoolS@sizey]%
    \foreach \x in {0, 1, ..., 5} \foreach \y in {0, 1, ..., 10}
    \coordinate (c\x\y) at (\x *0.12*\CoolS@sizex, \y *0.107*\CoolS@sizey);
    \draw [line width=\Cool@stroke] (c28)--(c26)--(c44)--(c42)--(c20)--(c02)--(c04)--(c15);
    \draw [line width=\Cool@stroke] (c22)--(c24)--(c06)--(c08)--(c210)--(c48)--(c46)--(c35);
\end{tikzpicture}}
\newcommand*\CoolS{%
  \mathchoice{\def\Cool@stroke{0.03*\f@size pt}\settoheight\CoolS@sizey{$\displaystyle\Gamma$}\settowidth\CoolS@sizex{$\displaystyle\Gamma$}\CoolS@inner}%
             {\def\Cool@stroke{0.03*\f@size pt}\settoheight\CoolS@sizey{$\textstyle\Gamma$}\settowidth\CoolS@sizex{$\displaystyle\Gamma$}\CoolS@inner}%
             {\mkern 1mu\def\Cool@stroke{0.03*\sf@size pt}\settoheight\CoolS@sizey{$\scriptstyle\Gamma$}\settowidth\CoolS@sizex{$\scriptstyle\Gamma$}\CoolS@inner}%
             {\mkern 1mu\def\Cool@stroke{0.03*\ssf@size pt}\settoheight\CoolS@sizey{$\scriptscriptstyle\Gamma$}\settowidth\CoolS@sizex{$\scriptscriptstyle\Gamma$}\CoolS@inner}%
}\makeatother

\begin{document}

\title{Spectroscopy From The Lattice: The Scalar Glueball}

\author{Ruair\'i Brett} 
\email[Corresponding author: ]{rbrett@gwu.edu}
\affiliation{
  Department of Physics, The George Washington University, Washington, DC 20052, USA}

\author{John Bulava}
\affiliation{%
  CP3-Origins, University of Southern Denmark, Campusvej 55, 5230 Odense M, Denmark}%

\author{Daniel Darvish} 
\affiliation{
  Department of Physics, Carnegie Mellon University, Pittsburgh, PA 15213, USA}

\author{Jacob Fallica} 
\affiliation{
  Department of Physics and Astronomy, University of Kentucky, Lexington, KY 40506, USA}

\author{Andrew~Hanlon} 
\affiliation{
  Helmholtz-Institut Mainz, Johannes Gutenberg-Universität, 55099 Mainz, Germany}

\author{Ben H\"orz} 
\affiliation{
  Nuclear Science Division, Lawrence Berkeley National Laboratory, Berkeley, CA 94720, USA}

\author{Colin Morningstar} 
\affiliation{
  Department of Physics, Carnegie Mellon University, Pittsburgh, PA 15213, USA}

\date{\today} 

\begin{abstract}
  Lattice calculations allow us to probe the low-lying, non-perturbative spectrum of QCD
  using first principles numerical methods. Here we present the low-lying spectrum
  in the scalar sector with vacuum quantum numbers including, in fully dynamical QCD for
  the first time, the mixing between glueball, q-qbar, and meson-meson operators.
\end{abstract}

\maketitle

\section{Introduction}
The three-gluon and four-gluon coupling terms in the QCD Lagrangian suggest the existence
of composite states consisting solely of gluons, called glueballs. Such states are of
great interest especially as they are distinct from the prototypical $\qbar q$ and
$qqq$ hadronic states predicted by constituent quark models. However, incontrovertible
experimental evidence for their existence remains elusive. There are several leading
candidates for the lightest scalar glueball, including the $f_0(1370)$, $f_0(1500)$, and
$f_0(1710)$ states, yet none have been unambiguously identified as a glueball
state~\cite{Crede2009}. To identify which of the three is most likely a glueball or
gluon-dominated state, model independent, first principles lattice calculations are
required.

The glueball spectrum in pure Yang-Mills gauge theory has been extensively mapped
out~\cite{Bali1993,Morningstar1999,Chen2006}. The lowest-lying scalar and tensor
glueballs have previously been studied in quenched QCD, but the quenched
approximation makes such studies unreliable.
For the scalar glueball, quenched
calculations yield a glueball mass in the range $1.5-1.7$~GeV.
More recent studies which have included the effects of sea quarks on
glueballs~\cite{Bali2000,Hart2002,Hart2006,Richards2010,Gregory2012,Sun2018}, largely
agree with one another and with the quenched calculations in the scalar, and tensor
sectors, but such studies have not included meson-meson operators.

As the candidate glueball states lie well above many hadron thresholds, the inclusion of
glueball, $\qbar q$, and meson-meson operators is crucial for making any definitive
conclusions about the nature, or even existence of such glueball states.
Furthermore, as these states in infinite-volume manifest as unstable resonances, forming
any infinite-volume conclusions will require the determination of coupled channel
infinite-volume scattering amplitudes from finite-volume energies.
Here we present the low-lying finite-volume spectrum from Ref.~\cite{Brett2019} in the
scalar sector with vacuum quantum numbers, where glueball, $\qbar q$, and meson-meson
operators have been included for the first time in lattice QCD.


\section{Analysis Details}
Temporal correlation functions involving glueball operators are notoriously difficult to
measure in lattice QCD, requiring prohibitively large computational resources to achieve
even modest statistical precision. In the scalar sector the signal-to-noise ratio for
such observables falls extremely rapidly with increasing separation between source and
sink, as the relevant interpolating operators have large vacuum expectation values.
This prohibits the lattice from being too large, as the magnitude of these vacuum
fluctuations will scale with the lattice volume. On the other hand, due to the large
masses of these states, lattice studies of glueballs require very fine temporal lattice
spacings so that a reliable signal can be measured. As both of these considerations have
a significant effect on the required computational resources, we employ an anisotropic
lattice that is spatially coarse and temporally fine~\cite{Morningstar1997}.

We use a single anisotropic ensemble of $N_f=2+1$ clover-improved Wilson fermions
with $m_\pi \approx 390$~MeV, generated by the Hadron Spectrum
collaboration~\cite{Edwards2008,Lin2009}. Various ensemble parameters are listed
in~\cref{tab:hadspecParams}. Throughout we will quote energies as dimensionless ratios
using a reference mass: $m_{\rm ref} = 2m_K$ where $m_K$ is the kaon mass.

\begin{table}
  \caption[Details of the anisotropic ensemble used in the scalar glueball study.]
          {Details of the anisotropic ensemble used in the scalar glueball study.
            The anisotropy $\xi=a_s/a_t$ has been determined by enforcing the relativistic
            dispersion relation for the pion, though the value is insensitive to the
            hadron used.}
  \label{tab:hadspecParams}
  \centering
  \begin{tabular*}{1.0\textwidth}{@{\extracolsep{\fill}}| c | c c c c c c |}
    \hline
    $(L/a_s)^3 \times (T/a_t)$ & $N_{\rm cfgs}$ & $a_s$ & $\xi_\pi$ &
    $a_t m_\pi$ & $a_t m_K$ & $m_\pi L$ \\ \hline
    $24^3 \times 128$ & $551$ & $0.12$~fm & $3.4464(71)$ &
    $0.06901(17)$ & $0.09689(15)$ & $5.7$ \\ \hline
  \end{tabular*}
\end{table}

\subsection{Correlation Matrix Analysis}
Finite-volume stationary state energies are extracted from the matrix of
temporal correlation functions,
${\cal C}_{ij}(t) = \vacexp{{\cal O}_i(t)\overline{\cal O}_j(0)}$,
for which all-to-all quark propagation is evaluated using the stochastic LapH
method~\cite{Morningstar2011}.
To extract the finite-volume energies $E_n$, and operator overlap factors
$Z_j^n \equiv \bra{0}{\cal O}_j\ket{n}$, we solve the generalized eigenvalue
problem~\cite{Luscher1990}
\begin{equation}
  C(t)v_n(t,\tau_0) = \lambda_n(t,\tau_0)C(\tau_0)v_n(t,\tau_0),
\end{equation}
where $C_{AB}(t) \equiv {\cal C}_{AA}(\tau_N)^{-1/2} {\cal C}_{AB}(t) {\cal C}_{BB}(\tau_N)^{-1/2}$
is the normalized correlation matrix, with normalization time $\tau_N$, and
$\tau_0$ is referred to as the metric time. To do this, we define the ``rotated''
correlation matrix by
\begin{equation} \label{eq:singlePivot}
  \widetilde{D}(t) \equiv U^\dagger C(\tau_0)^{-1/2} C(t) C(\tau_0)^{-1/2} U,
\end{equation}
where the matrix $U$ is formed using the eigenvectors of
$C(\tau_0)^{-1/2} C(\tau_D) C(\tau_0)^{-1/2}$, for a single choice of the metric and
diagonalization times $(\tau_0,\tau_D)$. By diagonalizing only for a single time
separation $\tau_D$, we avoid diagonalizing the correlation matrix for late times where
significantly increased statistical noise can lead to a significant bias in the final
results. The diagonal elements of $\widetilde{D}(t)$ can then be shown to tend to,
in the limit of large time separations,
$\lambda_n(t) \propto e^{-E_n t}$~\cite{Blossier2009}. Then we can use single- and
multi-exponential (to account for excited state contamination) fits to the diagonal
elements of $\widetilde{D}(t)$ to determine the energies $E_n$ and overlaps $|Z_j^n|^2$.

The basis of single- and two-hadron interpolating operators
used is constructed to overlap maximally with the states of interest as described in
Ref.~\cite{Morningstar2013}, including the so-called TrLapH scalar glueball operator
constructed using the eigenvalues of the covariant Laplacian:
\begin{equation}
  {\cal O}_G \equiv -\Tr[\Theta(\sigma_s^2+\covLap)\covLap].
\end{equation}
The operator basis is chosen so as to saturate the spectrum of single- and two-particle
stationary states below $\approx 2m_{\rm ref}$. Along with conventional isoscalar
$\qbar q$ single-hadron operators and a scalar glueball operator, we include $\pi\pi$,
$\eta\eta$, and $K\bar{K}$ two-hadron operators with various definite back-to-back momenta
for each of the allowed two-body decays in the sector.

As we are concerned with states that share quantum numbers with the vacuum, interpolating
operators designed to transform irreducibly in the at-rest $A_{1g}^+$ irrep (where $g$
and $+$ denote positive spatial parity and $G$-parity, respectively) are expected to have
non-zero vacuum expectation values (VEVs). These VEVs must be subtracted in order to
extract the signal of interest:
\begin{equation}
  C_{ij}(t) \to \vacexp{{\cal O}_i(t)\overline{\cal O}_j(0)} -
  \vacexp{{\cal O}_i} \vacexp{\overline{\cal O}_j}.
\end{equation}
For correlation functions featuring the scalar glueball operator, this subtraction
presents some additional difficulty. Even in the moderately sized volume employed here,
the magnitude of $\vacexp{{\cal O}_G}$ is very large, and statistically very noisy.
The large uncertainties that manifest on the inclusion of the scalar glueball operator in
the operator basis requires us to use quite aggressive noise reduction techniques in
order to reliably extract a signal.

\begin{figure}[t]
  \centering
  \includegraphics[width=0.33\textwidth]{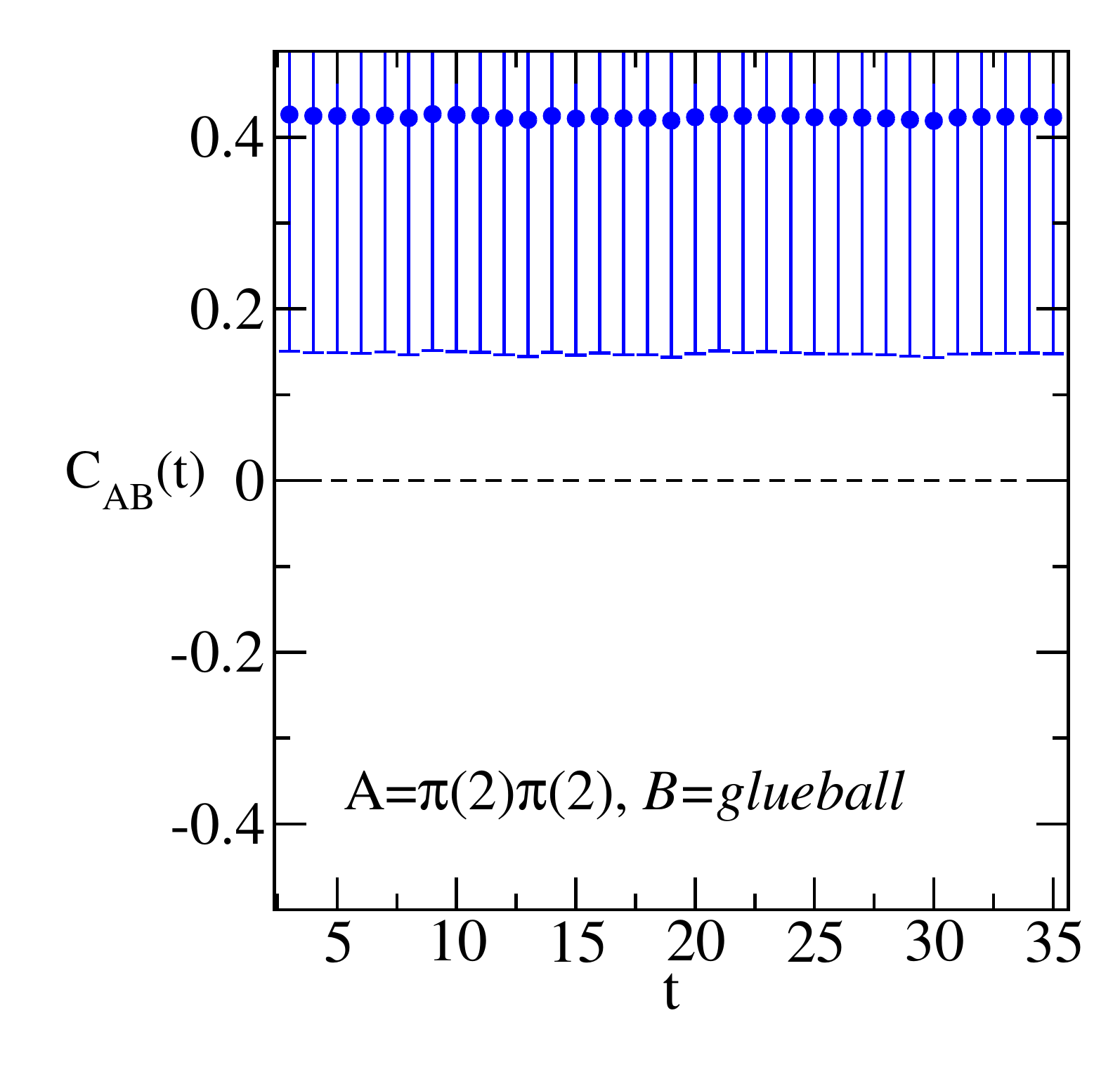}
  \includegraphics[width=0.33\textwidth]{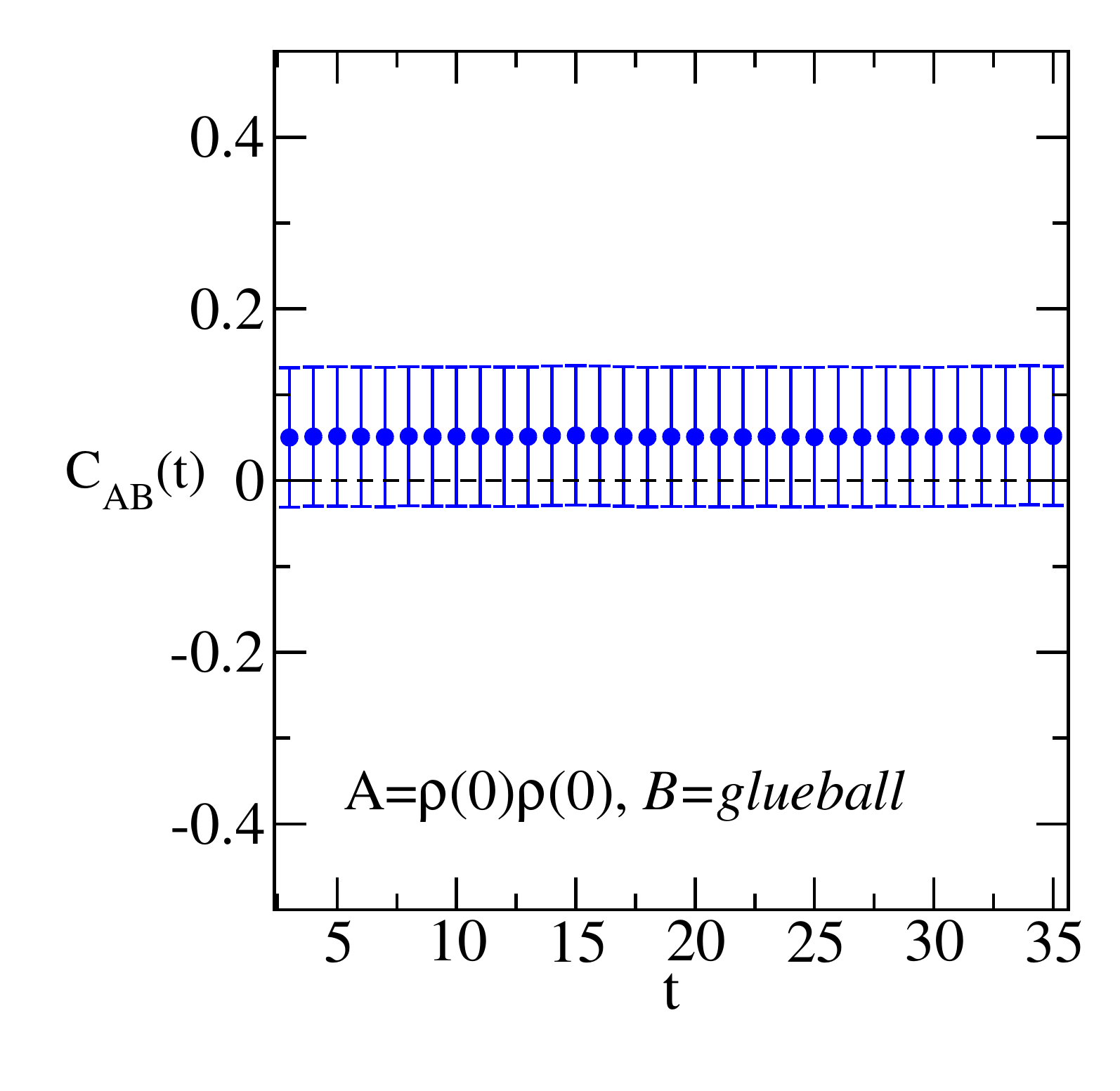}
  \includegraphics[width=0.33\textwidth]{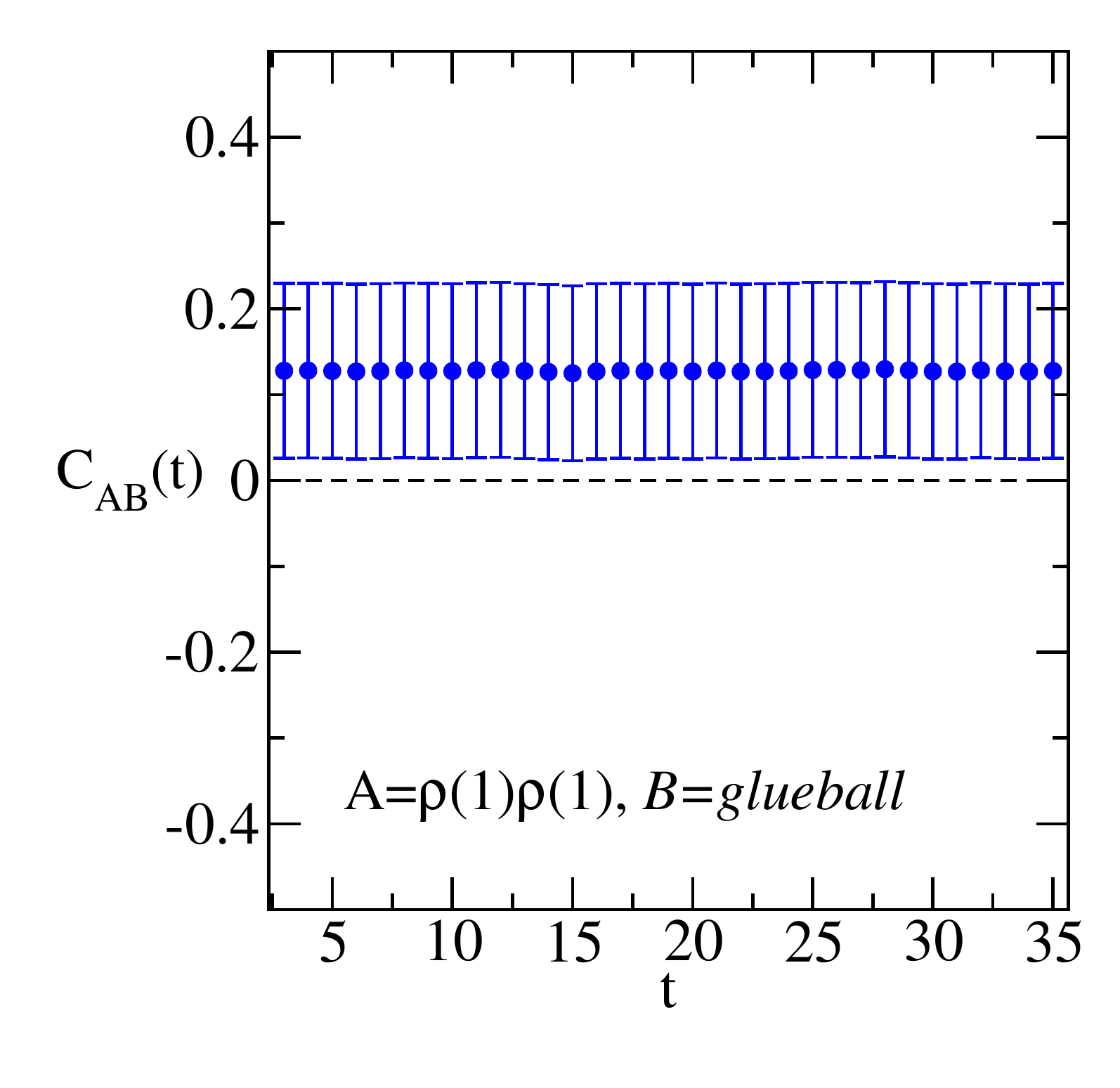}
  \caption{$\Im C_{AB}(t)$ for three normalized matrix elements including the scalar
    glueball operator. Each element has been normalized using
    $C_{AB}(t) \equiv {\cal C}_{AA}(\tau_N)^{-1/2} {\cal C}_{AB}(t) {\cal C}_{BB}(\tau_N)^{-1/2}$
    where ${\cal C}_{ij}(t) = \vacexp{{\cal O}_i(t)\overline{\cal O}_j(0)}$, and
    $\tau_N = 3$.}
  \label{fig:imCorr}
\end{figure}

Symmetry arguments based on the behaviour of our operators under time reversal, etc. tell
us that the correlation matrices here must be both real and symmetric.
In practice however, stochastic estimates of $\Im C_{ij}(t)$ will only be statistically
consistent with zero rather than exactly zero. This poses a problem when we include the
scalar glueball operator in our correlation matrix. See for example the matrix element
$\Im\vacexp{{\cal O}_G {\cal O}_{\pi(2)\pi(2)}}$ shown in~\cref{fig:imCorr}. 
The mean value is systematically shifted away from zero, hinting at the difficulty in
accurately estimating the large VEVs for the scalar glueball operator. We find that in
order to maintain a strictly positive definite correlation matrix when the basis includes
the scalar glueball operator, we must explicitly set the imaginary components of the
correlation matrix to be zero. We find that when the glueball operator is omitted, the
finite-volume spectrum extracted is unaffected by setting $\Im C_{ij} = 0$.

\begin{figure}[t]
  \centering
  \includegraphics[width=0.33\textwidth]{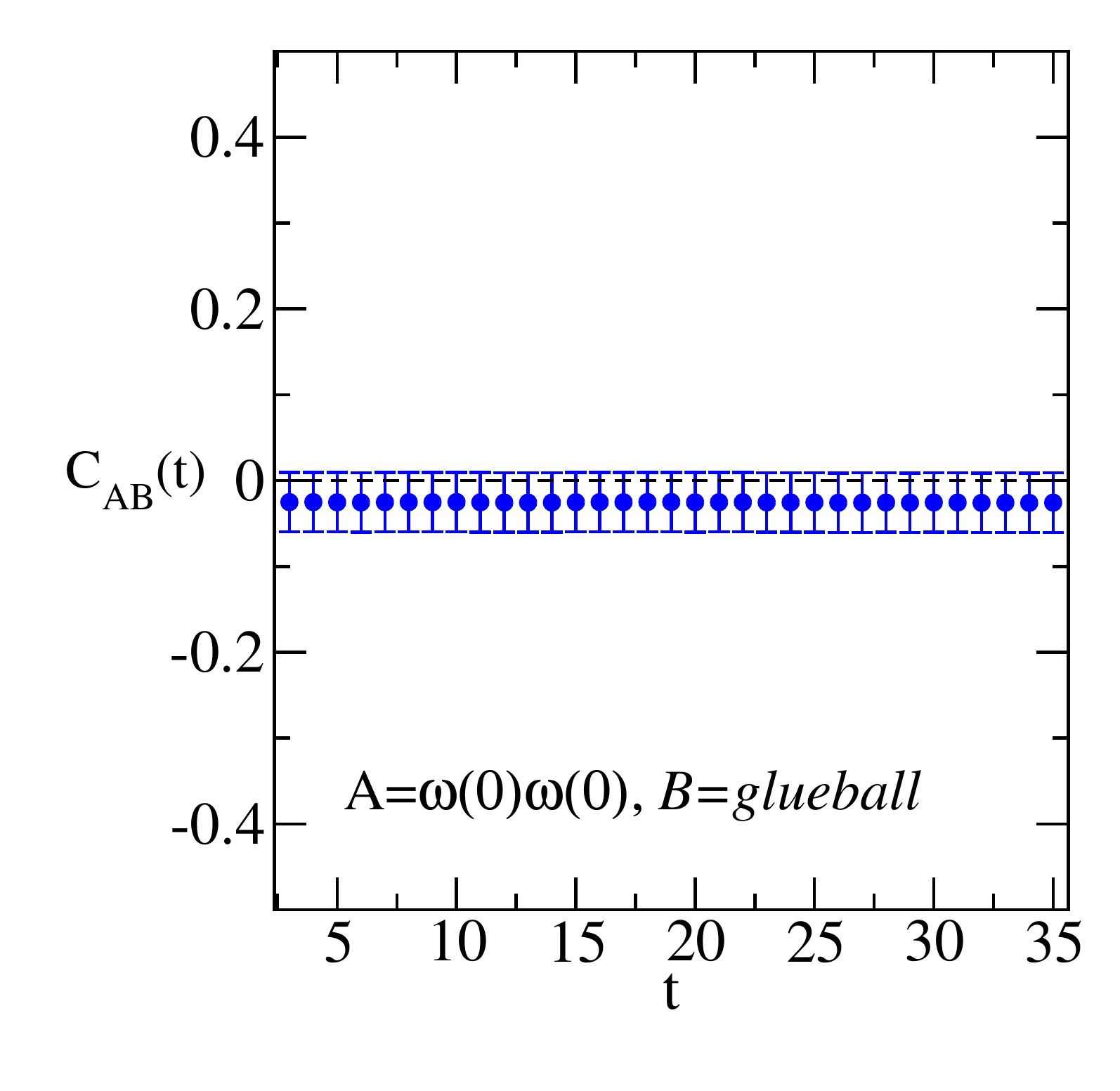}
  \includegraphics[width=0.33\textwidth]{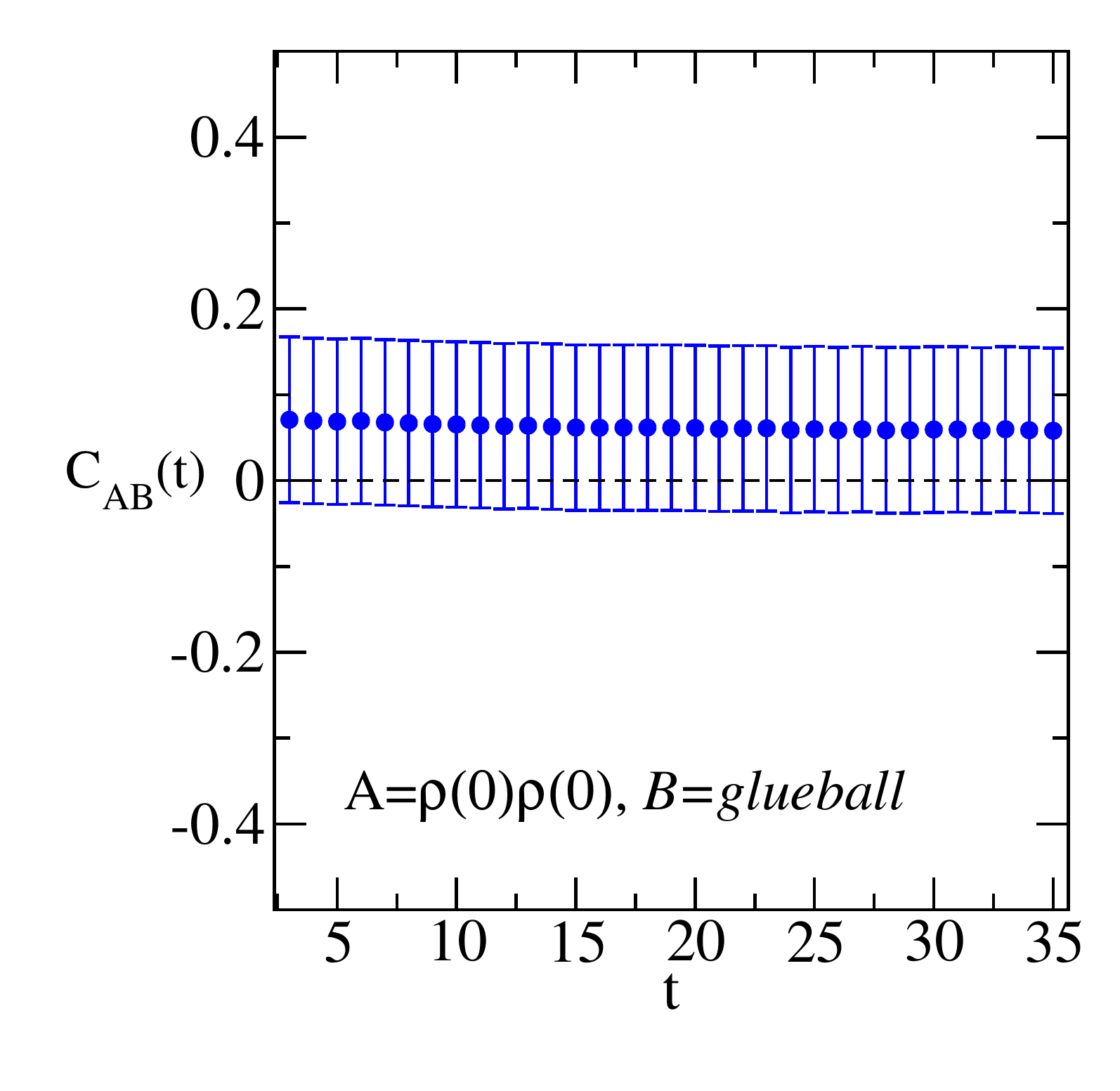}
  \includegraphics[width=0.33\textwidth]{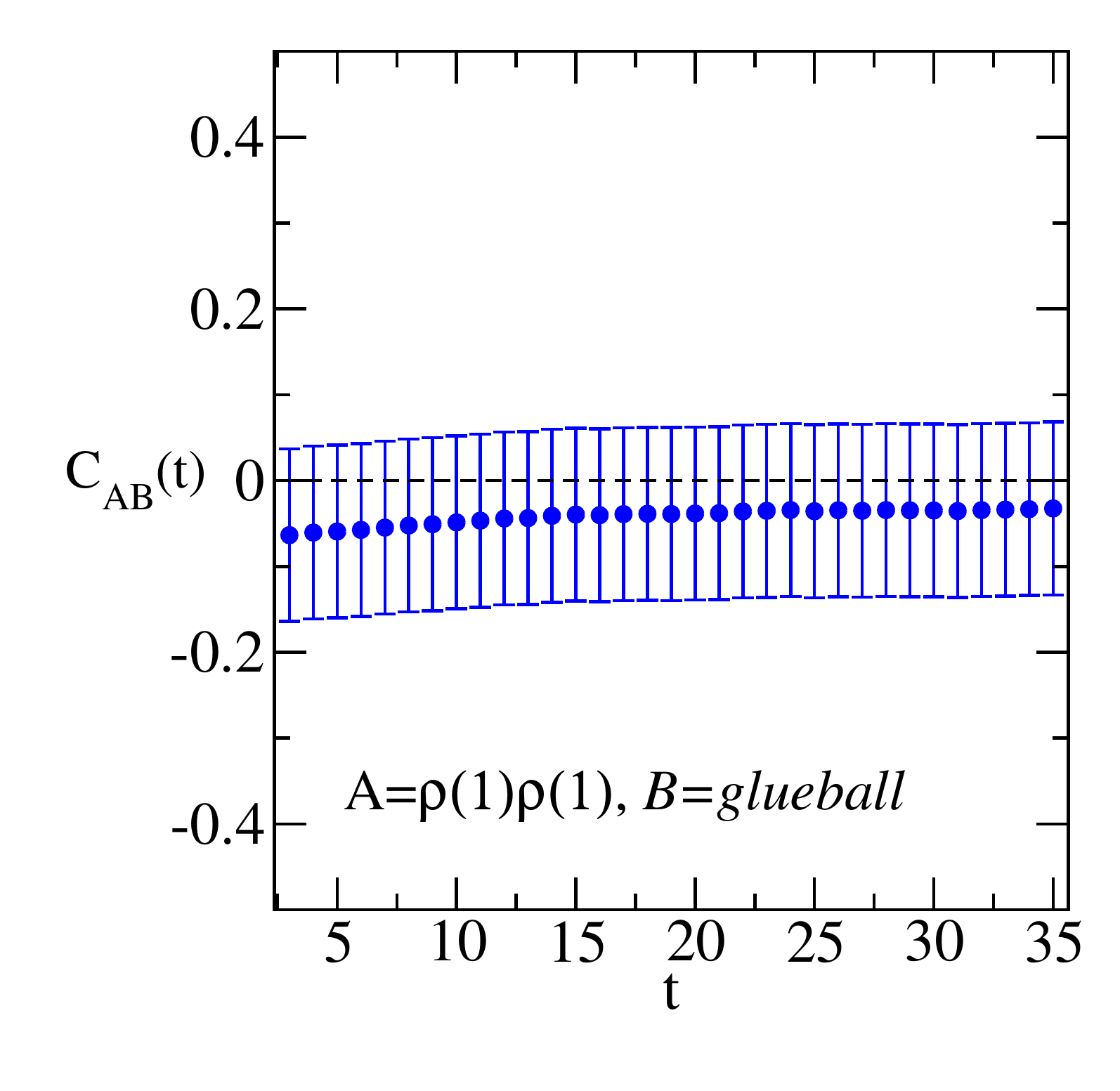}
  \caption{$\Re C_{AB}(t)$ for three normalized matrix elements including the scalar
    glueball operator and a vector-vector two-hadron operator. Each element has been
    normalized using
    $C_{AB}(t) \equiv {\cal C}_{AA}(\tau_N)^{-1/2} {\cal C}_{AB}(t) {\cal C}_{BB}(\tau_N)^{-1/2}$
    where ${\cal C}_{ij}(t) = \vacexp{{\cal O}_i(t)\overline{\cal O}_j(0)}$, and
    $\tau_N = 3$.}
  \label{fig:vectorGlue}
\end{figure}

The significant statistical noise present in the VEV-subtracted correlators presents
additional difficulty when we consider the ``single-pivot'' method
in~\cref{eq:singlePivot}. As we perform the diagonalization on only the full sample
estimate of the correlation matrix (\ie the matrix $U$), significant statistical noise
in the matrix elements can introduce a bias. Usually this is easily avoided by choosing
early diagonalization times for which statistical noise is minimized, however the
significant increase in noise, even at very early time separations, in the presence of
the glueball operator can have a drastic effect on the pivot. This is shown
in the $\vacexp{{\cal O}_G{\cal O}_{VV}}$ elements in~\cref{fig:vectorGlue}, where
${\cal O}_{VV}$ is a vector-vector two-hadron operator. We found that a significant bias
in the pivot could be mitigated by setting these elements that are statistically
consistent with zero to be exactly zero in our analysis.

\section{Spectrum Results}
\begin{figure}[t]
  \centering
  \includegraphics[width=0.6\textwidth]{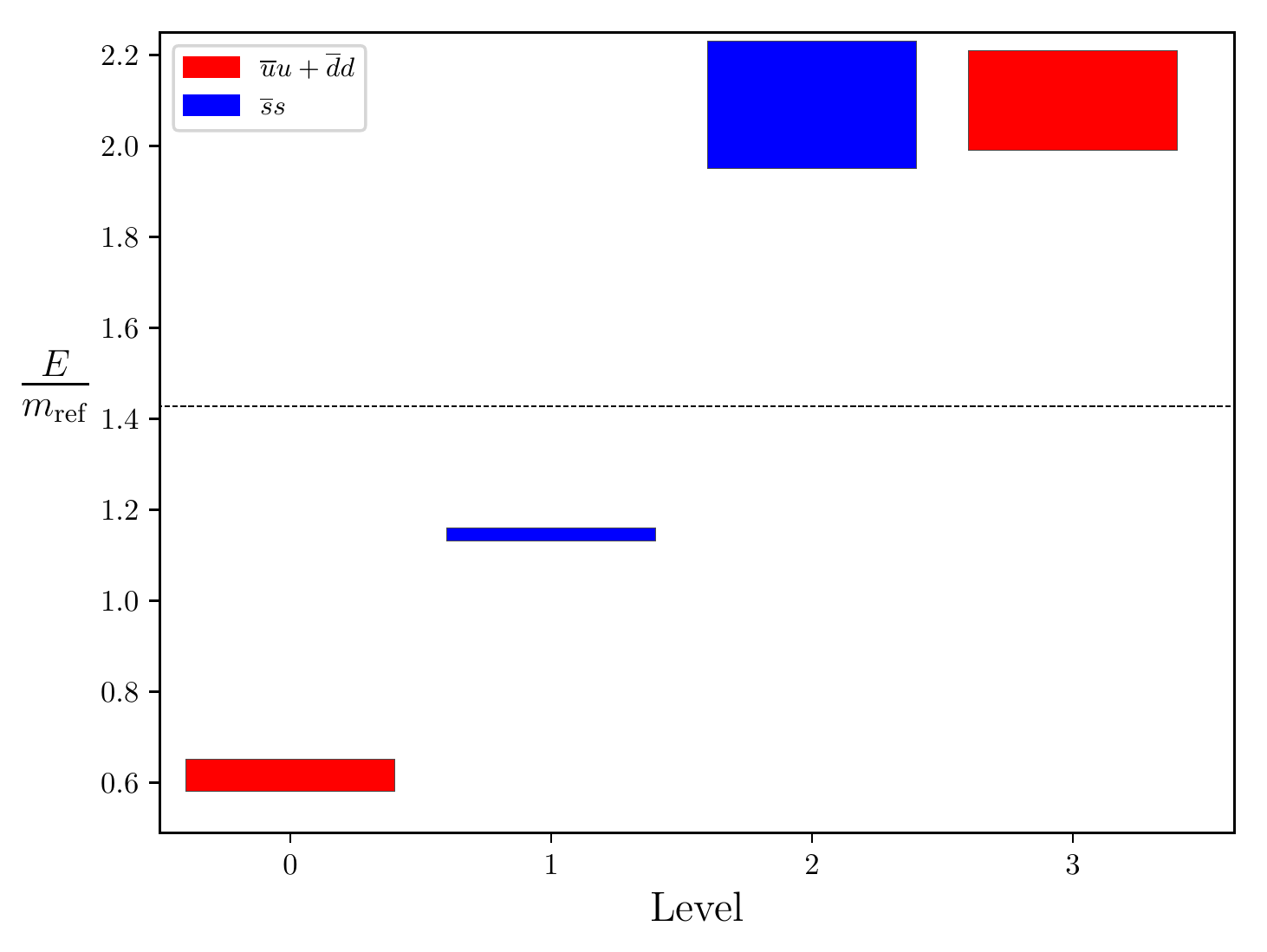}
  \caption[Flavoured staircase plot for the $A_{1g}^+$ channel using only $\qbar q$-type
    operators]
          {Finite-volume energies in the $I=0$, $S=0$, $A_{1g}^+$ channel for levels with
            significant overlap onto states produced only by quark-antiquark operators.
            A $4\times4$ correlation matrix including only $\qbar q$ operators is used
            to extract these levels
            . $1\sigma$ uncertainties are denoted by the
            box heights. Levels are coloured indicating the operator flavour type with
            maximal overlap onto that state. The horizontal dashed black line indicates
            the $4\pi$ threshold, and $m_{\rm ref} = 2m_K$. These energies do not change
            appreciably when other $\qbar q$ operators are included in a larger
            correlation matrix with meson-meson operators and the glueball operator.}
  \label{fig:qqbarStaircase}
\end{figure}

Our main goal is to discover if any finite-volume states below $2m_{\rm ref}$ in the
vacuum sector are \textit{missed} when no glueball operators are included. We assume that
our operators couple minimally to states involving three or more hadrons.
First we will consider the finite-volume spectrum determined using a
basis of interpolating operators \textit{excluding} the scalar glueball operator.
We begin by including a two-hadron (meson-meson) operator for each expected
non-interacting level, adding additional operators with various flavor, spin, etc.
structure until no new finite-volume levels are found below $\sim 2m_{\rm ref}$.
Single-hadron $\qbar q$ operators are chosen in a similar way, including one of each
isoscalar flavor structure: ($\ubar u + \dbar d$, $\sbar s$) with various spatial
displacements until no new states are seen in the energy region of interest.
We find only two such finite-volume states below $2m_{\rm ref}$ using $\qbar q$
operators, shown in~\cref{fig:qqbarStaircase}. Hence, we need only include two of these
operators in the final operator set, one of each flavor structure. This has also been
confirmed by including additional $\qbar q$ interpolating operators in the final
operator set and observing no deviation of the finite-volume spectrum below
$2m_{\rm ref}$.

\begin{figure}
  \centering
  \includegraphics[width=1.0\textwidth]{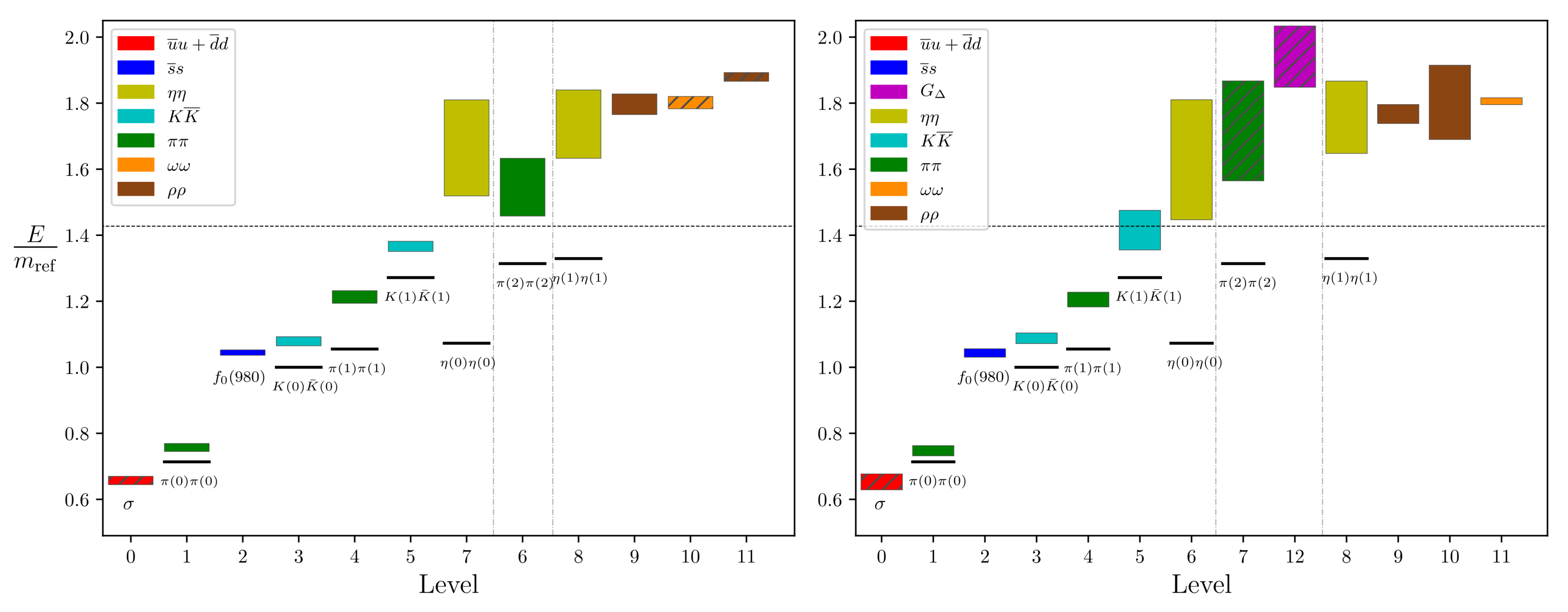}
  \caption{Finite-volume stationary state energies in the $I=0$, $S=0$, $A_{1g}^+$
            channel extracted using a $12\times12$ correlation matrix, \textit{excluding}
            the scalar glueball operator on the left, and using a $13\times13$ correlation
            matrix \textit{including} the scalar glueball operator on the right.
            $1\sigma$ uncertainties are denoted by the box
            heights. If a level is created predominantly by a single operator, the level
            is colored to indicate the flavor content of that operator. If a level is
            created predominantly by more than one operator, a hatched box is used to
            denote the presence of operator overlaps within $75\%$ of the maximum,
            indicating significant mixing. Level numbers indicate order in terms
            of increasing mean energy, but the levels have been rearranged horizontally
            to highlight the area of interest involving the glueball operator.
            Note that these finite-volume energies should \textit{not} be directly
            compared to the spectrum of experimental resonance states, in particular the
            two-hadron dominated levels. See text for further discussion. Short black
            lines indicate the non-interacting two-hadron levels, and the dashed
            horizontal black line indicates the $4\pi$ threshold, and
            $m_{\rm ref} = 2m_K$.}
  \label{fig:combStaircase}
\end{figure}

The finite-volume spectrum extracted using an operator basis \textit{excluding} the
glueball operator is shown on the left in~\cref{fig:combStaircase}. The significant
statistical noise present in many of the operators used here necessitates rather early
GEVP metric and diagonalization times of $(\tau_0, \tau_D) = (3,6)$, though we have used
various combinations of $\tau_0=3,4$, $\tau_D=4,5,6,7,8$ in order to ensure the spectrum
does not change. With these choices the correlation matrices remain well conditioned,
having condition numbers $<10$ at $\tau_0$ and $\tau_D$. We also ensure that the
off-diagonal elements of $\widetilde{D}(t)$ remain statistically consistent with zero
for $t>\tau_D$. We then \textit{include} the scalar glueball operator in the basis and
extract the finite-volume spectrum as above using $(\tau_0, \tau_D) = (3,6)$, shown on
the right in~\cref{fig:combStaircase}.

\begin{figure}
  \centering
  \includegraphics[width=1.0\textwidth]{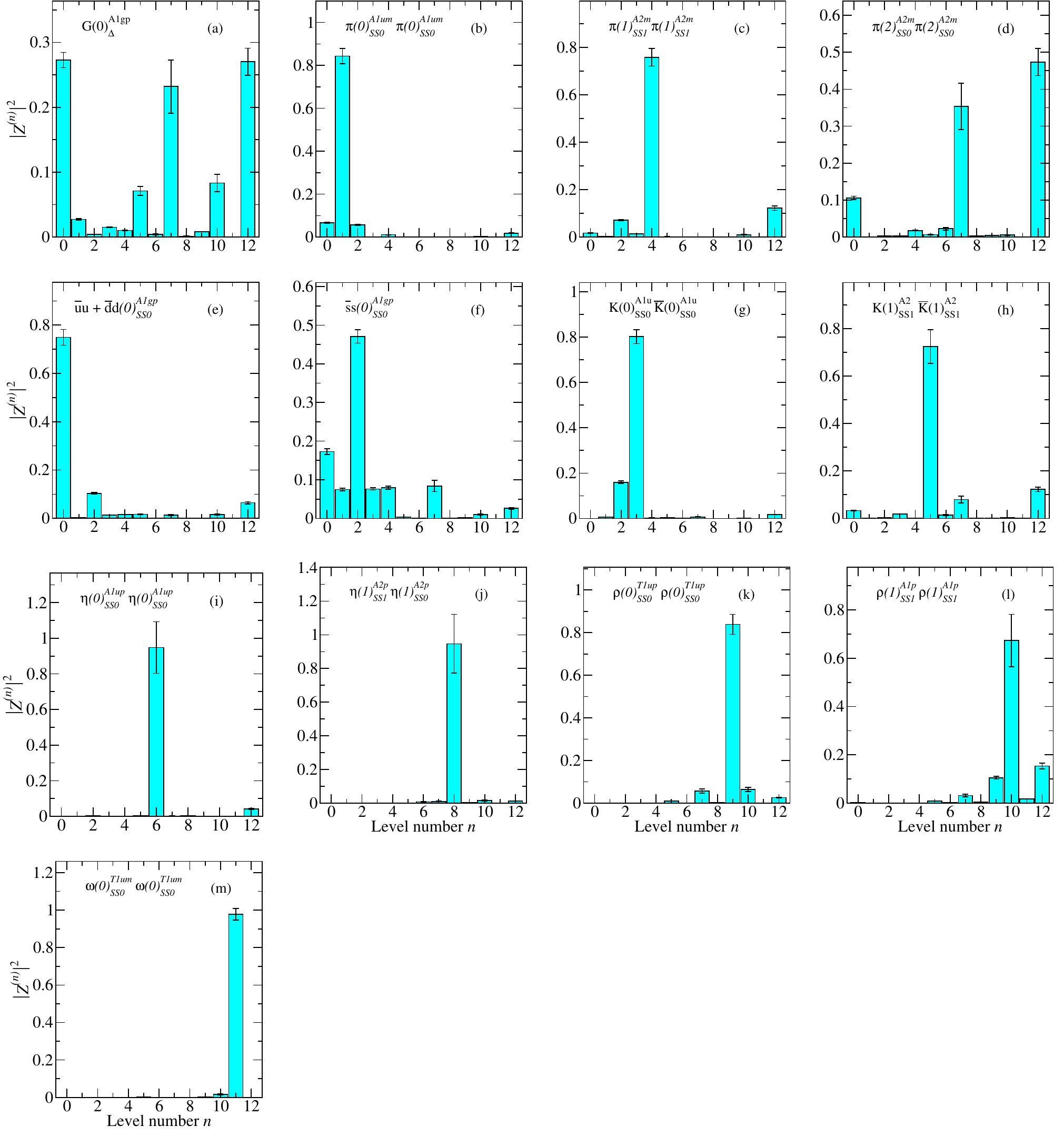}
  \caption{Overlaps $|Z^{(n)}|^2$ for the operators used in the $A_{1g}^+$ correlation
    matrix \textit{including} the scalar glueball operator.}
  \label{fig:glueballOverlaps}
\end{figure}

Looking first at the states below $4\pi$ in~\cref{fig:combStaircase}, indicated by the
horizontal dashed line, with the exception of some increased statistical noise, the
spectrum below $4\pi$ is insensitive to the addition of the glueball operator. The
overlap factors in~\cref{fig:glueballOverlaps} show minimal mixing in this region and so
level identification is relatively straightforward and is indicated by the colouring of
the energy levels.
As level 0 is predominantly created by the $(\overline{u}u+\overline{d}d)$ quark-antiquark
operator, along with the glueball operator, we can interpret this state as
the finite-volume counterpart of the $\sigma$ resonance. This is consistent with the
$\pi\pi$ scattering study of Ref.~\cite{Briceno2017} where a bound state $\sigma$ meson
is found below the $\pi\pi$ threshold.
Similarly, from figs. 4(b) and 4(f), levels 1 and 2 are created by the $\pi(0)\pi(0)$
and $s\overline{s}$ quark-antiquark operators, respectively, where the integers
indicate the square of the hadron momentum, in units of $2\pi/L$. As level 2 is
predominantly created by a $\qbar q$ interpolating operator, we identify level 2 as the
finite-volume counterpart of the $f_0(980)$ resonance, just above the $K\bar{K}$
threshold.

Above the $4\pi$ threshold, we can assess the effect that including the glueball operator
has on the finite-volume spectrum. Note that these levels in~\cref{fig:combStaircase}
have been reordered slightly. From ~\cref{fig:glueballOverlaps}(k-m) we can identify the
rightmost three levels as being predominantly created by the vector-vector
$\omega(0)\omega(0)$, $\rho(0)\rho(0)$, and $\rho(1)\rho(1)$ operators. With the exception
of an increase in statistical noise, these levels are largely unaffected by the inclusion
of the glueball operator. The same can be said for levels 6 and 8 (as numbered on the
right-hand side of~\cref{fig:combStaircase}), identifiable as being created dominantly by
the $\eta(0)\eta(0)$ and $\eta(1)\eta(1)$ type operators, respectively. Note however the
significant shift of these levels from the corresponding non-interacting values,
especially when compared to the shifts seen in the lower lying levels.

The remaining states, where the effect of the glueball operator is seen the most, are
highlighted by the vertical dashed lines. \Cref{fig:glueballOverlaps}(a) shows that the
glueball operator mainly creates levels 0, 7, and 12. Remarkably, a new state is
\textit{not} created near 1.5-1.7$m_{\rm ref}$. When the glueball operator is included,
there are two effects: the uncertainty in level 7 is greatly increased and
\textit{an additional state appears at a very high energy}. Based on
both~\cref{fig:glueballOverlaps}(d) and the overlap factors when the glueball operator is
excluded, we can identify level 7 as being dominantly created by $\pi(2)\pi(2)$.
When the glueball operator is included, it has significant overlap with this state.
More notable is that the additional state we extract with the enlarged operator basis
lies above all other extracted states in~\cref{fig:combStaircase}. This indicates that
we have saturated the spectrum of single and two-particle states in this region without a
glueball operator. Hence, we identify no finite-volume energy eigenstate predominantly
created by a scalar glueball operator below $\sim1.9m_{\rm ref}$. As this new energy
occurs above the region where our operator set is designed to create states, this level
appears most likely just as a consequence of the enlarged operator basis. We cannot
conclude that a pure glueball state has been created.

While these finite-volume results are insufficient to make any definitive
statements regarding the infinite-volume resonances in this channel, we can make some
qualitative comparisons to experiment. In finding only two $\qbar q$ dominated
states below $2m_{\rm ref}$, we have observed no clearly identifiable counterpart
finite-volume $\qbar q$ states to the $f_0(1370)$, $f_0(1500)$, or $f_0(1710)$ resonances
in this region. This \textit{suggests} that 
these resonances are molecular in nature rather than conventional $\qbar q$ or
pure glueball states.

\section{Conclusions}
We have presented here the first study of the low-lying spectrum in the scalar sector of
QCD with vacuum quantum numbers to include the mixing between $\qbar q$, two-hadron, and
glueball operators in fully dynamical lattice QCD. When a scalar glueball operator is
included in the operator basis, we observe an additional state lying near $2m_{\rm ref}$,
the upper limit we can study here. As the extracted spectrum, up to an increase in
statistical noise, is insensitive to the inclusion of the glueball operator, we have
concluded that the energies presented saturate the spectrum of single and two-particle
states in this region. Considering the leading experimental glueball candidates in this
region, the $f_0(1370)$, $f_0(1500)$, and $f_0(1710)$, we found no quark-antiquark
dominated levels identifiable as finite-volume counterpart to these states, suggesting
that these states are likely to be molecular in nature.

Our results reinforce, via the significant coupling of the glueball operator to the
$\pi(2)\pi(2)$ and $\sigma$ finite-volume states, the need for extensive operator bases
in a proper determination of the excited state spectrum in this sector of QCD.
To date, previous studies of glueballs in lattice QCD have included only glueball
interpolating operators. Forming definite infinite-volume conclusions about these states
will also require the determination of coupled-channel scattering amplitudes above the
$4\pi$ threshold. Hence, 3- and 4-hadron interpolating operators, along with a formalism
for extracting infinite-volume scattering information from finite-volume energies will be
required. Work in this direction is underway, with the spectrum of three-pion states with
maximal isospin reported recently in Ref.~\cite{Horz2019}, and a review of the current
state of amplitude extraction above the three particle threshold in
Ref.~\cite{Hansen2019}.

\begin{acknowledgments}
  This work was supported by the U.S. National Science Foundation under award PHY-1613449,
  and the John Peoples Jr. Research Fellowship at CMU. Computing resources were provided
  by the Extreme Science and Engineering Discovery Environment (XSEDE) under grant number
  TG-MCA07S017.
\end{acknowledgments}

\bibliography{refs}

\end{document}